\documentclass[submission, Phys]{SciPost}

\usepackage{amsfonts}
\usepackage{amssymb}
\usepackage{soul}

\begin{document}

\begin{center}{\Large \textbf{
  Finite-temperature critical behavior\\
  of long-range quantum Ising models
}}\end{center}

\begin{center}
E. Gonzalez Lazo\textsuperscript{1,2*},
M. Heyl\textsuperscript{3},
M. Dalmonte\textsuperscript{1,2},
A. Angelone\textsuperscript{1,2}
\end{center}

\begin{center}
{\bf 1} SISSA, via Bonomea 265, 34136 Trieste, Italy
\\
{\bf 2} The Abdus Salam International Centre for Theoretical Physics, Strada
  Costiera 11, 34151 Trieste, Italy
\\
{\bf 3} Max Planck Institute for the Physics of Complex Systems, Nöthnitzer
  Straße 38, Dresden 01187, Germany
\\
* \href{mailto:egonzale@sissa.it}{egonzale@sissa.it}
\end{center}

\begin{center}
\today
\end{center}


\section*{Abstract}

\textbf{
  We study the phase diagram and critical properties of quantum Ising chains
  with long-range ferromagnetic interactions decaying in a power-law fashion
  with exponent $\alpha$, in regimes of direct interest for current trapped ion
  experiments. Using large-scale path integral Monte Carlo simulations, we
  investigate both the ground-state and the nonzero-temperature regimes. We
  identify the phase boundary of the ferromagnetic phase and obtain accurate
  estimates for the ferromagnetic-paramagnetic transition temperatures. We
  further determine the critical exponents of the respective transitions. Our
  results are in agreement with existing predictions for interaction exponents
  $\alpha > 1$ up to small deviations in some critical exponents. We also
  address the elusive regime $\alpha < 1$, where we find that the universality
  class of both the ground-state and nonzero-temperature transition is
  consistent with the mean-field limit at $\alpha = 0$. Our work not only
  contributes to the understanding of the equilibrium properties of long-range
  interacting quantum Ising models, but can also be important for addressing
  fundamental dynamical aspects, such as issues concerning the open question of
  thermalization in such models.
}

\vspace{10pt}
\noindent\rule{\textwidth}{1pt}
\tableofcontents\thispagestyle{fancy}
\noindent\rule{\textwidth}{1pt}
\vspace{10pt}

\section{\label{sec:introduction}Introduction}

Systems featuring long-range interactions are central in condensed matter and
statistical physics, due to both their widespread presence in nature and the
wide range of characteristic physical phenomena they display, the latter often
being at odds with well-known predictions and results concerning short-range
models (see, e.g,~\cite{Campa2009} for a review).
Within the last decade, the interest in quantum long-range interacting models
has further surged due to the progress in manipulating and controlling these
systems at an unprecedented level ~\cite{Britton2012, Schau2012, Landig2016,
Monroe2019, Browaeys2020}.
Specifically, these experimental platforms naturally realize long-range quantum
Ising or Heisenberg models, with the possibility to engineer many-body
interaction potentials decaying proportionally to $d^{-\alpha}$ as a function
of distance $d$, ranging from van-der-Waals-like ($\alpha = 6$) and dipolar
interactions ($\alpha = 3$) in the context of Rydberg
atoms~\cite{Schau2012, Browaeys2020}, to Coulomb ($\alpha = 1$) and infinite-range
($\alpha = 0$) potentials for trapped ions~\cite{Britton2012, Monroe2019}.

Recent experiments in such long-range interacting models have mostly centered
on the investigation of inherent dynamical phenomena, such as many-body
localization~\cite{Smith2016}, discrete time
crystals~\cite{Zhang2017timecrystal, Choi2017},
prethermalization~\cite{Neyenhuis2017}, Kibble-Zurek
mechanism~\cite{Keesling2019, Scholl2020}, or dynamical quantum phase
transitions~\cite{Zhang2017transition, Jurcevic2017}.
Despite of recent progress~\cite{Fratus2016, Russomanno2020} one key question
has, however, remained open: especially in the limit of small interaction
exponents, it is not known whether these long-range systems follow the
fundamental principle of thermalization as expected for generic short-range
models.
In the first place, this obviously requires a thorough understanding of the
thermal properties of the system of interest, which have only been partially
explored even in paradigmatic Hamiltonians such as the one-dimensional
long-range quantum Ising model.

In particular, the ground-state properties of the latter in the case of
ferromagnetic (FM) interactions have been the focus of investigation via analytical
and renormalization group (RG) techniques~\cite{Dutta2001, Defenu2017, Maity2019}, as well
as linked-cluster expansions~\cite{Fey2016}, tensor network approaches and/or
density matrix RG~\cite{Zhu2018, Gabbrielli2019}, Monte Carlo
methods~\cite{Sperstad2012} and, very recently, Stochastic Series Expansion
(SSE) Monte Carlo~\cite{Koziol2021} investigation in the $\alpha > 1$ region,
demonstrating, e.g., that the critical behavior of the model belongs to the
mean-field and short-range universality class (UC) for $1 <\alpha < 5/3$ and
$\alpha \geq 3$, respectively.
The antiferromagnetic case has also been intensely studied via the use of
several approaches~\cite{Koffel2012, Fey2016, Sun2017, Rader2019,
Koziol2021}, with notable results including, among others, the demonstration
that the half-chain entanglement entropy displays area-law violations in the
intermediate regime $1 < \alpha < 2$~\cite{Koffel2012}. Similarly, $p$-wave superconductors with long-range
pairing~\cite{Vodola2015} have been shown to display exotic critical behavior, even if, due to the presence of
Jordan-Wigner strings, those models do in general differ from Ising chains with similarly decaying interactions.
Considerable effort has also been dedicated to the theoretical investigation of
the dynamical properties of this type of model~\cite{Pappalardi2018,
Defenu2018, Lerose2019-1, Lerose2019-2, Piccitto2019-1, Piccitto2019-2,
Khasseh2020}.

Oppositely with respect to the zero-temperature case, the finite-temperature
regime is still poorly understood.
Indeed, the latter has been predicted by general theoretical
arguments~\cite{Sachdev} to belong to the universality class of the
corresponding classical long-range Ising model, with quantum effects not
changing this description at the qualitative level.
While this picture has been essentially confirmed for the case $\alpha = 2$ by
SSE studies~\cite{Humeniuk2020}, the latter demonstrated, in the proximity of
the ground-state critical point, the presence of considerable finite-size
effects induced by strong quantum fluctuations, which all but prevent
observation of the expected classical regime even at very large system sizes.

In the light of the experimental realizations of these models discussed above,
investigating the thermal critical behavior of these Hamiltonians remains
therefore of great importance, in order to determine the role and strength of
the quantum effects in perturbing the predicted classical picture.
Furthermore, (numerically) exact analysis of the finite-temperature regime is
essential to determine non-universal details such as, e.g., the position of
thermal critical points, which are influenced in a key way by quantum effects,
and whose knowledge is crucial for laboratory realizations.
Such a study is of especially great interest in the extremely long-ranged
regime $0 < \alpha < 1$, which, to our knowledge, has not been the object of
this kind of investigation, and (as mentioned above) is directly realizable in
trapped-ions setups.

\begin{figure}[]
  \centering

  \includegraphics[width=\columnwidth]{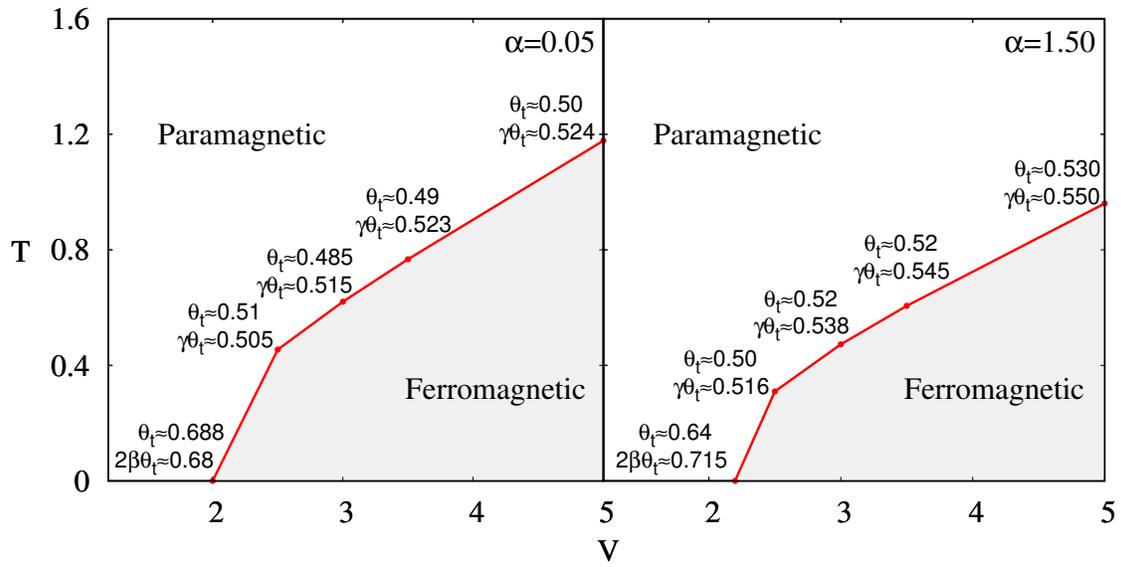}

  \caption{Calculated phase diagram of the long-range transverse-field Ising
  model in eq.~\eqref{eq:hamil}, displaying the ground-state and
  finite-temperature phase boundary and critical exponents obtained using
  finite-size scaling techniques. Panels (a) and (b) correspond to $\alpha =
  0.05$ and $\alpha = 1.50$, respectively. Here, $T$ is the system temperature
  in units of the Boltzmann constant, and $V$ is the interaction strength in
  units of the transverse field (see below). The displayed results for the
  effective thermal exponent and its product with the magnetization and
  susceptibility critical exponent are those obtained via data collapse (see
  below).}

  \label{fig:PhaseDiagram}
\end{figure}

In this work, we study both the ground-state and finite-temperature phase
diagram of the long-range FM quantum Ising model in one spatial
dimension, by means of numerically exact, large-scale Path Integral Monte Carlo
simulations.
We perform our calculations for two representative values of $\alpha$: namely,
we choose $\alpha = 0.05$ and $\alpha = 1.50$, within the extremely long-range
region $\alpha < 1$ and intermediate region $1 < \alpha < 2$, respectively.
We employ a wide variety of well-known finite-size scaling techniques to
determine the position (i.e., the critical points) and critical exponents of
both the ground-state and finite-temperature paramagnetic (PM)-FM
transitions displayed by the model, obtaining the phase diagram displayed in
Fig.~\ref{fig:PhaseDiagram}.

We determine the critical points and critical exponents for the ground-state
FM-PM transition. Our results for critical point positions
and correlation length critical exponents are in agreement with existing
predictions in the literature where the latter are available (i.e., $\alpha =
1.50$), while we encounter relatively small ($\sim 7\%$) deviations with
respect to our estimate for the magnetization critical exponent.
We then obtain accurate results for the position of the critical points in the
finite-temperature regime for several values of the interaction strength.
Concomitantly, our estimated correlation length critical exponents at $\alpha =
1.50$ essentially confirm the theoretical prediction of no qualitative
deviations from the classical universality class due to quantum fluctuations,
while discrepancies (up to $10 \%$ in the strongly interacting region) appear
in the susceptibility critical exponent.

The structure of the paper is the following.
Sec.~\ref{sec:Model and methods} introduces the Hamiltonian, the numerical
technique employed for its study, and the finite-size scaling approaches we
employed to analyze its critical behavior.
Sec.~\ref{sec:Results} discusses our obtained results on the critical
behavior of the model.
Finally, in Sec.~\ref{sec:conclusions} we outline the conclusions of our work
and offer an outlook for future direction of research.

\section{\label{sec:Model and methods}Model and methods}

\subsection{Hamiltonian and known results}

The model analyzed in this work is described by the Hamiltonian
\begin{equation}
  H = - \frac{V}{K(L)} \sum_{i < j} \frac{S_i^z S_j^z}{r_{ij}^{\alpha}} - h
  \sum_i S_i^x,
  \label{eq:hamil}
\end{equation}
where $V > 0$ is the interaction strength, $i,j$ run over the sites $1, \ldots,
L$ of a one-dimensional lattice with periodic boundary conditions, $r_{ij}$ is
the distance between sites $i$ and $j$, $S_i^z$ ($S_i^x$) is the component
along $z$ ($x$) of the spin-1/2 operator acting on site $i$, and $K(L) \equiv
(L - 1)^{-1} \sum_{i \neq j} r_{ij}^{- \alpha}$ is the Ka\'c renormalization
factor.
The latter ensures the existence of a proper thermodynamic limit in the regime
$\alpha \leq 1$, while for $\alpha > 1$ it amounts to a rescaling of the
interaction strength, and does not change the universal features of the
critical behavior of the model.
We remark that the presence of this renormalization factor is directly related
to how interactions with $\alpha < 3$ are engineered in trapped ions
experiments. The latter exploit coupling between the ions and collective modes
of the ion chain (phonons), mediated via a single laser shined over the full
sample.
Increasing the number of ions while keeping the lattice spacing constant
naturally leads to a reduced coupling strength, that translates into the fact
that the energy of the full system is still extensive - as reflected by Ka\'c
normalization.
In the following, periodic boundary conditions are taken into account following
the minimum-image convention, and $h = 1$ will be taken as unit of energy.

For very small interaction strength $V$, the ground state of the system in the
thermodynamic limit is a paramagnet, characterized by a vanishing value of the
magnetization along the $z$ direction $|m_z| \equiv L^{-1} |\sum_i S_i^z|$.
On the contrary, for $V \gg 1$ the system is in a FM phase,
displaying a finite $|m_z|$.
The existence of a finite-$V$ phase transition connecting these two states can
be proven via analytical arguments (see, e.g., \cite{Dutta2001}); its UC
depends strongly on the value of the decay parameter $\alpha$.
Indeed, the $\alpha = 0$ case, also referred to as \textit{Lipkin-Meshkov-Glick
model} \cite{Lipkin1965}, can be described in an exact fashion at the
mean-field level~\cite{Botet1983}, and the PM-FM
transition has been proven to belong to the Gaussian UC in the $1 < \alpha <
5/3$ region.
In contrast, in the regime $\alpha \geq 3$, the critical point belongs to the
short-range UC (i.e., the one of the FM-PM transition in
the nearest-neighbor limit $\alpha \to \infty$).

In the finite-temperature regime, generic scaling arguments \cite{Sachdev}
predict that the model should display the same critical behavior as its
classical (i.e., $h = 0$) counterpart, due to the finiteness of the system size
in the imaginary time dimension (see below).
The critical behavior of the classical model has been studied via both
analytical (see, e.g., \cite{Dyson1969}), RG (see, e.g., \cite{Fisher1972}) and
numerical techniques (see, e.g., \cite{Luijten1997}) in the $\alpha > 1$
regime. Here, the system displays a second-order FM-PM
thermal phase transition for $1 < \alpha < 2$, with the region $1 < \alpha <
3/2$ belonging to the mean-field regime, while in the point $\alpha = 2$ the
model undergoes a finite-temperature transition of the BKT type, and the
short-range regime is reached (i.e., no finite-temperature transition takes
place) for $\alpha > 2$.

\subsection{Numerical techniques and finite-size scaling}

We perform our investigation of the Hamiltonian in eq.~\eqref{eq:hamil} via
Path Integral Monte Carlo (PIMC)~\cite{Sandvik1997}, a numerically exact
technique for the study of unfrustrated systems of bosons and quantum spins.
In this approach, one maps the features of a quantum model of interest to those
of an equivalent, higher-dimensional classical one, which is then studied via
Metropolis Monte Carlo simulations.
The quantum-to-classical mapping described above maps the partition function of
the extended transverse-field Ising model in eq.~\eqref{eq:hamil} into the one
of an anisotropic extended Ising model on a rectangular lattice, via a
procedure known as \textit{Suzuki-Trotter breakup}.
Here, in addition to the original spatial dimension, one also considers a
discretized and periodic one, known as \textit{imaginary time}, which extends
in the interval $\left[ 0, \beta \right]$, where $\beta = 1/T$ is the inverse
system temperature in units of the Boltzmann constant.
The number of sites $M$ along this direction (also known as \textit{slices}) is
a free parameter which affects the accuracy of the mapping: indeed, the latter
is exact up to $O(\beta/M)$ corrections, which vanish in the limit $M \to
\infty$.

In the spatial direction, the extended Ising model resulting from the mapping
displays the same FM long-range interactions present in the
spin-spin term of the model in eq.~\eqref{eq:hamil}, while spin-spin couplings
are nearest-neighbor in the imaginary time direction.
Our PIMC algorithm combines conventional Wolff cluster updates \cite{Wolff1989}
in imaginary time with efficient long-range cluster updates \cite{Luijten1997}
in the spatial direction.
The choice of these two state-of-the-art techniques allow to accurately analyze
large system sizes (up to $L = 8192$ sites) at low enough temperatures (down to
$\beta = 1024$) to reach the ground state regime.
The Suzuki-Trotter corrections mentioned above are kept into account by
performing simulations with increasing number of slices (up to $M = 65536$),
until a value $M = M^*$ is found such that the corresponding values of the
observables of interest were determined to be identical, within statistical
error, to those obtained for $M = 2M^*$.
The same protocol (with $\beta$ in the place of $M$) is adopted to ensure the
$T \to 0$ limit is reached in the investigation of the ground state regime.

The PIMC algorithm gives us direct access to observables commuting with the
$S_i^z$ operators, including the integer powers of $|m_z|$. This allows us to
compute quantities such as the Binder cumulant
\begin{equation}
  U = \frac{1}{2} \left[ 3 - \frac{\langle m_z^4 \rangle}{\langle m_z^2
  \rangle^2} \right],
  \label{eq:Binder}
\end{equation}
where $\langle \ldots \rangle$ stands for statistical averaging, which is
expected to converge to $1$ ($0$) in a FM (PM) phase
\cite{Sandvik2010}.
We also compute the ``classical'' susceptibility
\begin{equation}
  \chi = \beta L \left( \langle m_z^2 \rangle - \langle |m_z| \rangle^2 \right),
  \label{eq:Susdef}
\end{equation}
which, in proximity of a finite-temperature critical point of a quantum model,
approximates well the exact functional form of the magnetic susceptibility
\cite{Humeniuk2020}.

In order to extract reliable information on the critical behavior of the model
in the thermodynamic limit, we exploit the well known finite-size scaling (FSS)
theory \cite{Sandvik2010}.
In this framework, scaling relations of various quantities in terms of the
correlation length $\xi$, which diverges when approaching a critical point, are
exploited to obtain finite-size information by noting that in a finite system
$\xi$ will saturate to a value $O(L)$, where $L$ is the system size.
Features such as the position of the critical point or the critical exponents,
on which the original scaling relations depended, can then be directly
extracted via numerical fits as a function of $L$.
In the following section, when discussing the fitting procedures to obtain such
quantities, we will offer detailed formulae regarding FSS predictions for
observables such as $U$ and $\chi$.

\section{\label{sec:Results}Results}

We investigate the critical properties of the model in eq.~\eqref{eq:hamil} in
the ground-state and finite-temperature regime for $\alpha = 0.05$ and $\alpha
= 1.50$.

\subsection{Ground-state critical behavior}

The first step in our analysis is the determination of the
PM-FM critical point $V_c$ in the ground-state regime,
which we accomplish by fitting to our numerical data for the Binder cumulant
$U$ its expected FSS behavior.
The Binder cumulant curves $U(V)$ for system sizes $L$ and, e.g., $2L$ are
expected to cross at size-dependent points $V = V_U(L)$, which will follow (to
the leading order) the FSS scaling \cite{Angelini2014, Koziol2021}
\begin{equation}    
  V_U(L) = V_c \left( 1 + aL^{- \omega - \theta_t} \right),
  \label{eq:scalingBC1}
\end{equation}
where $V_c$ is the critical point, and the \textit{effective thermal exponent}
$\theta_t$ is linked to the correlation length critical exponent $\nu$.

In the ground-state regime $\nu^{-1} = \theta_t$ outside of the mean-field
region; conversely, when the latter is entered, corrections to the leading
scaling behavior can be taken into account \cite{Koziol2021} via the
generalized expression $\nu^{-1} = \left( d_{\mathrm{uc}}(\alpha)/d \right)
\theta_t$, where $d$ is the dimensionality and $d_{\mathrm{uc}}(\alpha) = 3
(\alpha-1)/2$ is the upper critical dimension for the value of $\alpha$ of
interest.

\begin{figure}[]
  \includegraphics[width=\columnwidth]{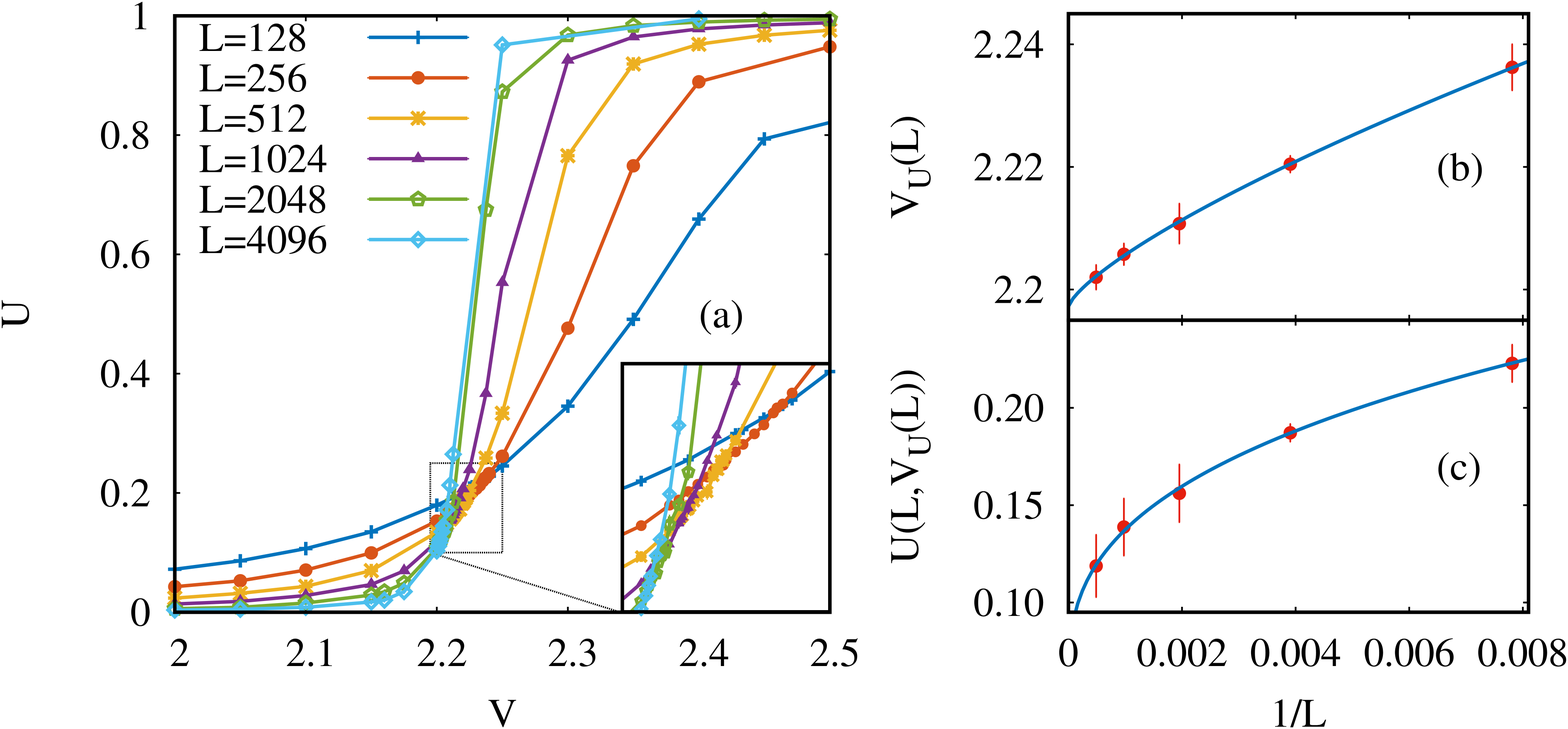}
  \caption{Binder cumulant scaling in the ground state regime (in all panels,
  $\alpha = 1.50$). Panel (a): Binder cumulant curves as a function of $V$ for
  different system sizes. Solid lines are a guide to the eye. Inset:
  magnification of the curve crossing region. Panel (b): computed crossing
  positions $V_U(L)$ between the Binder cumulant curves at system sizes $L$ and
  $2L$. The continuous line is a numerical fit to the expected FSS behavior in
  eq.~\eqref{eq:scalingBC1}. Panel (c): computed values of the Binder cumulant
  at the crossing points $V_U(L)$ between system sizes $L$ and $2L$. The
  continuous line is a numerical fit to the predicted FSS behavior in
  eq.~\eqref{eq:scalingBC2}.}
  \label{fig:CrossingsA1.5.eps}
\end{figure}

Comparison of eq.~\eqref{eq:scalingBC1} with the predicted leading-order FSS
behavior for the \textit{value} of the Binder cumulant at the $V_U(L)$s,
\begin{equation}
  U(L, V_U(L)) = b + cL^{- \omega},
  \label{eq:scalingBC2}
\end{equation}
allows us to obtain estimates for $V_c$ and $\theta_t$, by fitting our computed
results for the crossing features [see Fig.~\ref{fig:CrossingsA1.5.eps}(a)]
with the functional forms above.

Fig.~\ref{fig:CrossingsA1.5.eps}(b-c) display examples of the FSS fitting
procedures mentioned above; the obtained values of the critical point and of
the effective thermal exponent $\theta_t$ are listed in
Table~\ref{tab:tableGroundPhase}.

\begin{table}[h!]
  \centering

  \begin{tabular}{c|cc|cc|c}
	  $\alpha$ & $V_c$ (BC) & $V_c$ (DC) & $\theta_t$ (BC) & $\theta_t$ (DC) & $2 \beta_m \theta_t$ (DC) \\ \hline
    $0.05$ & $1.9997(4)$ & $1.9999$ & $0.50(7)$ & $0.688$ & $0.68$ \\
    $1.50$ & $2.1972(7)$ & $2.1981$ & $0.39(6)$ & $0.64$ & $0.715$
  \end{tabular}

  \caption{\label{tab:tableGroundPhase} Values of $V_c$, $\theta_t$, and
  $\beta_m$ (see text) associated to the ground state
  paramagnetic-ferromagnetic transition, computed via FSS analysis of the
  Binder cumulant crossings (BC) and via data collapse of the squared
  magnetization $m_z^2$ (DC).}
\end{table}

In order to gain more insight into the ground-state critical behavior of the
model, we perform a data collapse analysis by directly exploiting the FSS
predictions for the behavior of the squared magnetization close to a critical
point \cite{Sandvik2010, Koziol2021},
\begin{equation}
  m_z^2 \sim L^{-2 \beta_m \theta_t} \cdot f \left[ L^{+ \theta_t} \left( V_c -
  V \right) \right]    \qquad V \gtrsim V_c,
  \label{eq:mzScal}
\end{equation}
where $\beta_m$ is the magnetization critical exponent, up to corrections of
higher order in $1/L$.
This scaling law implies that the rescaled magnetization curves $y^{m}_L
\equiv m^2_z(L) L^{+ 2 \beta_m \theta_t}$ for different system sizes should
coincide if plotted as a function of $x^V_L \equiv \left( V_c - V \right)
L^{\theta_t}$.
We perform a high-order polynomial fit of $y^m_L$ as a function of $x^V_L$ in a
window around the critical point $x^V_L = 0$ for a wide range of candidate
values of $V_c$, $\theta_t$ and $\beta_m$, choosing as our final estimates for
these quantities the values which resulted in the fit with the lowest
chi-square value.
While it is hard to assign a rigorous error bar to the results of a
data collapse analysis, we estimate the order of magnitude of the error on our
results by performing the same fits in a considerably larger (i.e, containing
of the order of double the number of points) window around the critical point,
and taking the difference between the optimal values of $V_c$, $\theta_t$, and
$\beta_m$ for the two windows as the order of their numerical uncertainty.

\begin{figure}[]
  \includegraphics[width=\columnwidth]{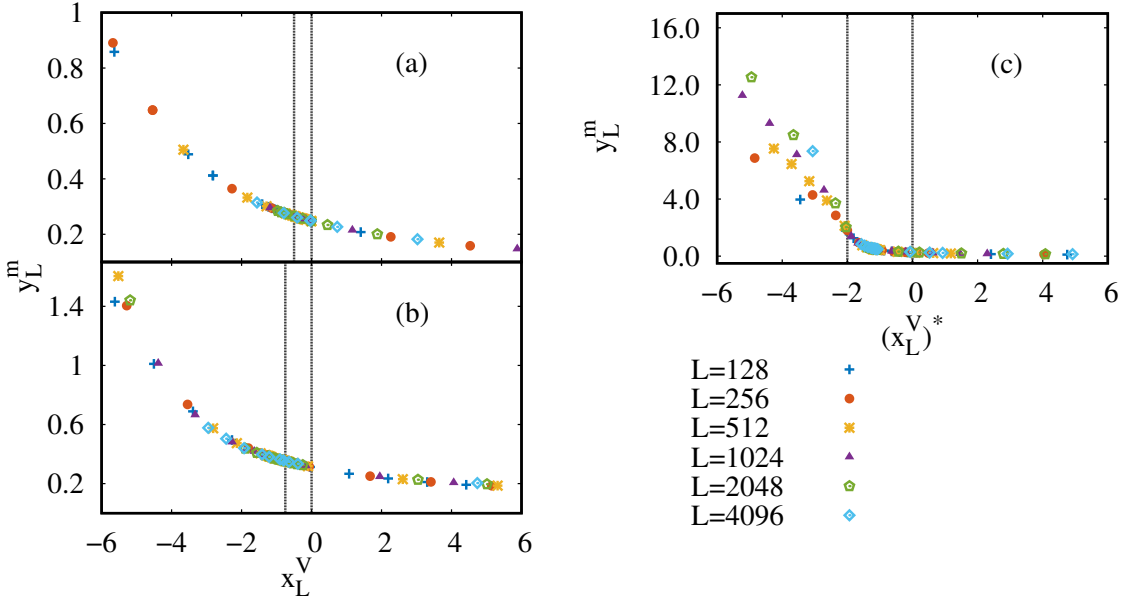}
  \caption{Panel (a):  data collapse of the rescaled squared magnetization
  $y^m_L$ as a function of the rescaled interaction strength $x^V_L$ for
  $\alpha = 0.05$. Panel (b): same as panel (a) for $\alpha = 1.50$. Panel
  (c): same as panel (b), where the data collapse rescaling is performed on
  the Ka\'c-factor-free rescaled interaction (see text). In all panels, the
  black dashed lines enclose the interval of the independent variable within
  which the data collapse scaling fit has been performed.}
  \label{fig:GroundCollapse}
\end{figure}

Our collapsed data is displayed in Fig.~\ref{fig:GroundCollapse}(a-b);
the obtained estimates for $V_c$, $\theta_t$ and $\beta_m$ are listed in
Table~\ref{tab:tableGroundPhase}.
We note that the data collapse behavior takes place over a fairly wide range
of values of the rescaled order parameter $x^V_L$, despite relatively narrow
fitting windows for the scaling behavior in eq.~\eqref{eq:mzScal} (the
intervals between dashed lines in Fig.~\ref{fig:GroundCollapse}).
This highlights the faithfulness of the data collapse scaling description of
our numerical data, which translates to highly reliable estimates of the
critical properties of the system.

Examination of our results points out i) the remarkable agreement of the
critical point estimates obtained via the Binder cumulant FSS and the
data collapse, and ii) conversely, the incompatibility between the two
estimates for the effective thermal exponent $\theta_t$.
Due to the arguments mentioned above, we believe the data collapse estimates
for the critical features to be more reliable in this regard.

For $\alpha = 1.50$, we find agreement for $\theta_t$ and deviations of the
order of 7\% for $2 \beta_m \theta_t$ from the independent SSE predictions in
Ref.~\cite{Koziol2021} which, in our notation, are $\theta_t \simeq 2 \beta_m
\theta_t \simeq 0.667$.
We also find good agreement with the estimate $V_c \simeq 0.42$ (in our
notation) given in \cite{Koziol2021} for the position of the ground-state
critical point, by performing a data collapse where the rescaled interaction
$x_L^V$ is replaced by $\left( x_L^V \right)^* \equiv L^{+ \theta_t} \left( V_c
- V/K(L) \right)$ (the rescaling is required since the Ka\'c correction factor
is not employed in \cite{Koziol2021}).
The resulting data collapse [see Fig.~\ref{fig:GroundCollapse}(c)] yields
optimal values $\theta_t \simeq 0.64$, $2 \beta_m \theta_t \simeq 0.76$, and $V_c
\simeq 0.42$.
For $\alpha = 0.05$, our estimates for $\theta_t$ and $2 \beta_m \theta_t$ are
compatible (up to deviations of the order of 3\% in $\theta_t$) with the ones
corresponding to the $\alpha = 0$ mean-field critical behavior, i.e., $\theta_t
= 2 \beta_m \theta_t = 2/3$~\cite{Botet1983}.

\subsection{Finite-temperature critical behavior}

Once the boundary of the ground-state FM phase is determined, we
investigate whether or not FM order survives for $T > 0$, and more
in general the details of the critical behavior of the model in this regime.
To this end, we perform finite-temperature calculations for fixed values of $V$
belonging to the FM phase in the ground state regime.
We apply the FSS framework to quantities such as the Binder cumulant and the
susceptibility, computed as a function of $T$, to estimate features of the
temperature-driven critical behavior.

\begin{figure}[]
\includegraphics[width=\columnwidth]{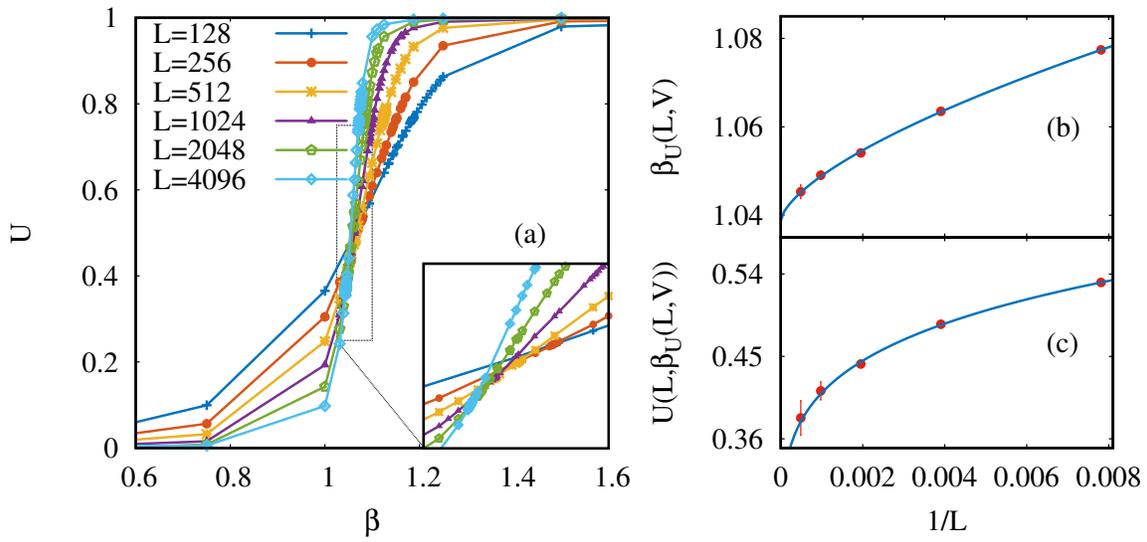}
\caption{Binder cumulant scaling in the finite-temperature regime (in all
  panels, $\alpha = 1.50$ and $V = 5.0$). Panel (a): Binder cumulant curves as
  a function of $\beta$ for different system sizes. Solid lines are a guide to
  the eye. Inset: magnification of the curve crossing region. Panel (b):
  computed crossing positions $\beta_U(L,V)$ between the Binder cumulant curves
  at system sizes $L$ and $2L$. The continuous line is a numerical fit to the
  expected FSS behavior in eq.~\eqref{eq:scalingBC1}. Panel (c): computed
  values of the Binder cumulant at the crossing points $\beta_U(L,V)$  between
  system sizes $L$ and $2L$. The continuous line is a numerical fit to the
  predicted FSS behavior in eq.~\eqref{eq:scalingBC2}.}
  \label{fig:CrossingsA1.5V5.0.eps}
\end{figure}

Indeed, our results for the Binder cumulant as a function of $\beta$ at fixed
$V$ and different system sizes immediately confirm the presence of a
finite-temperature phase transition, as pointed out by the appearance of the
crossing behavior discussed above [see Fig.~\ref{fig:CrossingsA1.5V5.0.eps}(a)]
at size-dependent points $\beta_U(L,V)$.
We determine the $V$-dependent critical temperatures $\beta_c(V)$ and the
associated $\theta_t(V)$ via fitting of the FSS relations in
eqs.~\eqref{eq:scalingBC1}-\eqref{eq:scalingBC2} to our computed crossing
features, with the thermal critical points $\beta_c$ and $\beta$ taking the
role of $V_c$ and $V$, respectively.
If the hypothesis of essentially classical critical behavior for the
finite-temperature quantum model holds (as we argue below) one may
link~\cite{Flores-Sola2015} $\theta_t$ to the correlation length critical
exponent $\nu$ via the relation $\nu^{-1} = \left(
d_{\mathrm{uc}}^{\mathrm{class}}(\alpha) / d \right) \theta_t$, where
$d_{\mathrm{uc}}^{\mathrm{class}}(\alpha) = 2 (\alpha -1)$ is the classical upper
critical dimension.

\begin{figure}[]
  \includegraphics[width=\columnwidth]{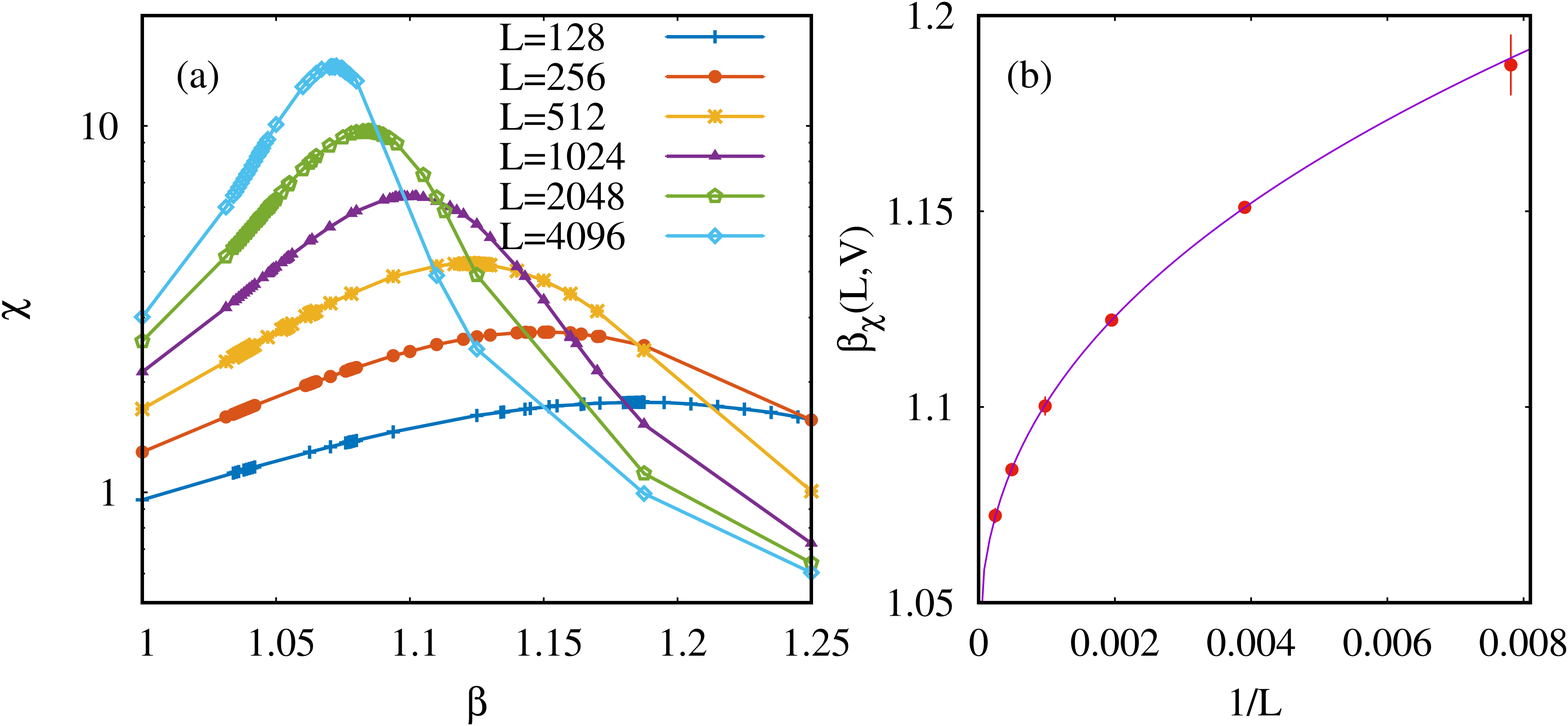}
  \caption{FSS analysis of the magnetic susceptibility in the
  finite-temperature regime (in all panels, $\alpha = 1.50$ and $V = 5.0$).
  Panel (a): susceptibility curves as a function of $\beta$ for different
  system sizes. Solid lines are a guide to the eye. Panel (b): finite-size peak
  positions $\beta_\chi(L)$. The continuous line is a numerical fit to the
  expected FSS behavior in eq.~\eqref{eq:SusScaling}.}
  \label{fig:SUSA1.5V5.0.eps}
\end{figure}

\begin{figure*}[h!]
  \includegraphics[width=0.9\textwidth]{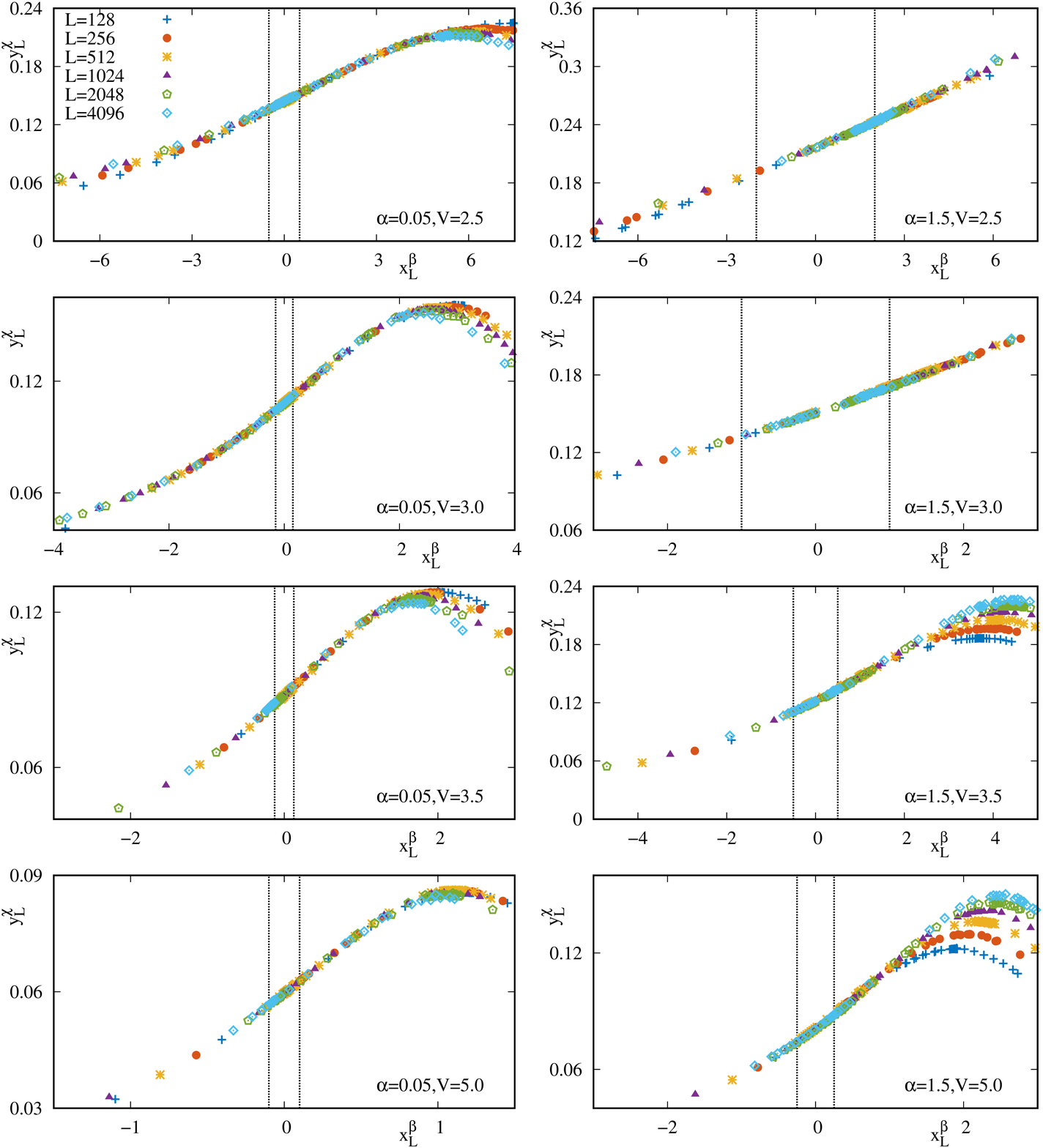}
  \caption{Data collapse of the rescaled magnetic susceptibility $y^{\chi}_L$
  as a function of the rescaled order parameter $x^{\beta}_L$ for the values of
  $\alpha$ and $V$ studied in this work. The black dashed lines enclose the
  interval of $x^{\beta}_L$ within which the data collapse scaling fit has
  been performed.}
  \label{fig:AllCollapse.eps}
\end{figure*}

Examples of this analysis are displayed in
Fig.~\ref{fig:CrossingsA1.5V5.0.eps}(b-c): the obtained critical parameters are
listed in Table~\ref{tab:tableFiniteTemperaturePhase}. 
We remark here that our application of this approach encountered in some cases
strong difficulties due to significant finite-size effects in proximity of the
$\beta_c(V,L)$.
In particular, the relatively large numerical uncertainties on the values of
the Binder cumulant in this region led to the necessity to perform conservative
estimates of the finite-size crossing points.
In turn, this prevented us in some cases from obtaining meaningful (i.e., with
small enough error bars) estimates for $\theta_t$.

In order to obtain an independent estimation of our quantities of interest, we
investigate the finite-temperature behavior of the magnetic susceptibility for
the same values of $V$ selected in our Binder cumulant analysis.
At finite system size and fixed interaction strength, $\chi$ is expected to
display peaks at size-dependent temperatures $\beta_{\chi}(L,V)$; the FSS
framework predicts for the latter \cite{Sandvik2010, Koziol2021} the leading
scaling behavior
\begin{equation}
  \beta_{\chi}(L,V) = \beta_c + f L^{- \theta_t}
  \label{eq:SusScaling}
\end{equation}
as a function of the system size.

Our numerical data confirm the expected behavior of $\chi$ [see
Fig.~\ref{fig:SUSA1.5V5.0.eps}(a)].
Fitting the FSS functional form in eq.~\eqref{eq:SusScaling} to the computed
peak positions [see Fig.~\ref{fig:SUSA1.5V5.0.eps}(b) for an example] allows us
to directly estimate the critical temperatures and effective thermal exponents
as a function of the interaction strength (see
Table~\ref{tab:tableFiniteTemperaturePhase} for a list of results).

While also requiring conservative estimates (and therefore large error bars) for
the peak positions, due to strong finite-size effects, we found the
susceptibility-based approach to be much less sensitive to this issue than the
Binder cumulant FSS discussed above.
In particular, we encountered problematic results only for $V = 2.5$, for both
values of $\alpha$ considered in this work, where our estimates were strongly
dependent on the set of system sizes considered in the fitting procedure (the
reported results correspond to the fits with all sizes considered).

We finally analyze the critical properties of the model by performing a
data collapse analysis for the behavior of the magnetic susceptibility close
to the finite-temperature critical points \cite{Sandvik2010, Koziol2021,
Luijten1997},
\begin{equation}
  \chi \sim L^{+ \gamma \theta_t} \cdot f \left[ L^{+ \theta_t} \left( \beta_c
  - \beta \right) \right] \qquad \beta \sim \beta_c,
   \label{eq:SusScal}
\end{equation}
where $\gamma$ is the susceptibility critical exponent, up to corrections of
higher order in $1/L$. The analysis follows the same protocol outlined in our
discussion of the ground-state regime, with the rescaled dependent and
independent variables here being $y_L^{\chi} \equiv \chi(L) L^{- \gamma
\theta_t}$ and $x^{\beta}_L \equiv \left( \beta_c - \beta \right)
L^{\theta_t}$, respectively.

\begin{table*}
  \begin{tabular}{cc|ccc|ccc|c}
    & & \multicolumn{3}{c|}{$\beta_c$} & \multicolumn{3}{c|}{$\theta_t$} & $\gamma \theta_t$ \\ \hline
                    & $V$ & $U$ & $\chi$ & $\chi_{dc}$ & $U$ & $\chi$ & $\chi_{dc}$ & $\chi_{dc}$ \\ \hline
    $\alpha = 0.05$ & $V = 2.5$ & $2.2007(4)$ & $2.23(1)$ & $2.20$ & / & $0.72(4)^*$ & $0.51$ & $0.505$ \\
  	                & $V = 3.0$ & $1.6120(7)$ & $1.61(1)$ & $1.612$ & / & $0.54(3)$ & $0.485$ & $0.515$ \\
                  	& $V = 3.5$ & $1.299(1)$ & $1.303(3)$ & $1.303$ & / & $0.54(2)$ & $0.49$ & $0.523$ \\
                    & $V = 5.0$ & $0.8474(2)^*$ & $0.844(2)$ & $0.8491$ & $0.5(1)$ & $0.47(2)$ & $0.50$ & $0.524$ \\ \hline
    $\alpha = 1.50$ & $V = 2.5$ & $3.21(1)$ & $3.351(9)$ & $3.229$ & $0.49(7)$ & $0.75(1)^*$ & $0.50$ & $0.516$ \\
                    & $V = 3.0$ & $2.109(1)^*$ & $2.12(1)$ & $2.115$ & $0.50(2)$ & $0.48(3)$ & $0.52$ & $0.538$ \\
                    & $V = 3.5$ & $1.647(6)$ & $1.646(5)$ & $1.650$ & $0.5(2)$ & $0.46(2)$ & $0.52$ & $0.545$ \\
                    & $V = 5.0$ & $1.039(1)$ & $1.035(1)$ & $1.041$ & $0.44(7)$ & $0.41(1)$ & $0.530$ & $0.550$ \\
  \end{tabular}

  \caption{\label{tab:tableFiniteTemperaturePhase} Summary of the computed
  estimates for $\beta_c$, $\theta_t$, and $\gamma \theta_t$ (see text) for the
  finite-temperature transitions at our investigated values of $\alpha$ and
  $V$. Our results are categorized according to the methodology employed to
  derive them: namely, FSS of the Binder cumulant crossings ($U$), FSS of the
  magnetic susceptibility peak position ($\chi$), and data collapse of the
  susceptibility ($\chi_{dc}$). Estimates marked with an asterisk ($*$) did not
  converge with respect to the choice of minimum size to be included in the
  fitting procedure.}
\end{table*}

Fig.~\ref{fig:AllCollapse.eps} displays our collapsed data for all the
values of $\alpha$ and $V$ investigated in this work; the corresponding optimal
(in the sense discussed above) results for $\beta_c$, $\theta_t$ and $\gamma$
are displayed in Table~\ref{tab:tableFiniteTemperaturePhase}.
As in the ground-state regime, we observe that the parameter range in which
the data collapse scaling \textit{ansatz} is respected noticeably exceeds our
fitting window (and vastly so, in most cases), highlighting the accuracy of
this approach in describing the critical behavior of the model.
Furthermore, this protocol does not require the estimation of size-dependent
features, sush as the curve crossings for the Binder cumulant, or the peak
position for the susceptibility, allowing us to obtain much more reliable and
systematics-free results.
We also note that high degree of accuracy with which the scaling law in
eq.~\eqref{eq:SusScaling} can be applied to describe the behavior of the
''classical'' susceptibility in eq.~\ref{eq:Susdef} is a strong indication of
the goodness of the latter as an approximation for the complete functional form
of the magnetic susceptibility.

A direct analysis of the results for the critical exponents listed in
Table~\ref{tab:tableFiniteTemperaturePhase} shows that our estimates obtained
via FSS of the Binder cumulant crossings, where meaningful in the sense
discussed above, are consistent within error bar with the ones obtained via
susceptibility data collapse.
Concomitantly, in some points we observe differences (which remain consistently
small, except for the point $\alpha = 1.50, V = 5.00$) between the latter and
the results of the susceptibility peak position FSS for the values of $V$ in
which the latter have converged with respect to the system sizes employed in
the fitting procedure.
In the points where this did not happen, the $\theta_t$ result from the
susceptibility peak position fit decreased, shifting towards the data-collapse
results, when smaller sizes were discarded.

According to the arguments mentioned in Sec.~\ref{sec:Model and methods}, the
universality class of the $T > 0$ FM-PM transition should
be the same of the corresponding transition in the classical counterpart of
model eq.~\eqref{eq:hamil}. For $\alpha = 1.50$, the classical Hamiltonian is
in the mean-field regime, and RG predictions, confirmed by classical Monte
Carlo calculations \cite{Luijten1997}, yield the estimates $\theta_t = \gamma
\theta_t = 1/2$.
Direct comparison with our most representative and reliable results in
Table~\ref{tab:tableFiniteTemperaturePhase} (i.e., the one obtained via
data collapse of the magnetic susceptibility) shows that our estimates for
$\theta_t$ are in essential agreement with the classical prediction (with
deviations outside of the estimated order of magnitude of the error only
appearing for $V = 5.0$).
Compatibility between our estimate and the theoretical predictions, even for $V
= 5.0$, is confirmed by the results obtained via FSS of the Binder cumulant,
while the susceptibility FSS estimates, where converged, show appreciable
deviations only for $V = 5.0$.
Conversely, our estimates for $\gamma \theta_t$ show relatively consistent
deviations (up to the order of 10\%), which increase with the interaction
strength.

These differences with the predicted results may be in principle due to several
causes, including i) the ``classical'' approximation employed for the study of
the susceptibility in our analysis, or ii) genuine quantum effects which
introduce deviations with respect to the predicted classical behavior.
However, we find it unlikely that either (i) and/or (ii) may be the dominant
physical mechanism underlying the observed deviations, since both effects are
essentially quantum in nature, and are expected to become weaker for larger
values of $V$, where in contrast our results are more at odds with the
classically predicted values.
Indeed, for higher interaction strengths quantum effects are expected to
weaken, due to both the larger value of $V$ (in comparison to the transverse
field $h$) and the higher temperature at which the critical region is located.
This consideration leads us to the conclusion that despite these deviations
(which may be caused by finite-size effects, or by higher-order corrections)
the critical behavior of the model in this regime follows the classical UC.

As in the ground-state case, we find essential compatibility with the
(classical) mean-field exponents at $\alpha = 0$; in particular, we match the
predicted values~\cite{Botet1983} $\theta_t = \gamma \theta_t = 1/2$ up to
relatively small deviations (of up to $2.5\%$) for the latter quantity, which
also become larger in the strongly interacting regime, and are therefore likely
not due to genuine quantum effects as argued above.

\section{\label{sec:conclusions}Conclusions and outlook}

We study the ground-state and finite-temperature phase diagram and critical
behavior of the long-range quantum Ising model in one spatial dimension, for
values of the interaction exponent parameter of direct interest
for current experiments in trapped ion setups.
We perform numerically exact, large-scale PIMC simulations within both the
extremely long-range region and intermediate long-range regime, respectively,
employing a wide variety of finite-size scaling techniques to determine the
location (i.e., the critical points) and critical exponents of both the
ground-state and finite-temperature phase transitions displayed by the model.

We determine transition points and critical exponents for the ground-state
FM-PM transition. We find essential agreement with
existing predictions for these quantities, where available (up to small
deviations for the value of the magnetization critical exponent), and
compatibility of our extremely-long-range results with the fully-connected
universal properties.
We then accurately estimate the position of the critical points in the
finite-temperature regime for several values of the interaction strength.
Here, our estimated critical exponents in the intermediate-long-range region
essentially confirm the theoretical prediction of classical universality.
In particular, in the intermediate long-range regime our estimated correlation
length critical exponent is fully consistent with the classical predictions,
while the susceptibility exponent displays deviations at most up to the order of $10
\%$.
Similarly, in the extremely long-range region we find compatibility with the
(classical) mean-field universality class up to deviations of the order of $2.5
\%$ in the value of the correlation length critical exponent. For future works, it would be interesting to verify
if some of these findings also apply to long-range $p$-wave superconductors~\cite{Vodola2015}, that, while described by
free theories, could still display some of the phenomenology we discuss.

Beyond exploring the equilibrium phase diagram and the nature of critical
points, our work is also directly relevant for another open question appearing
in the context of quantum Hamiltonians with long-ranged interactions.
This concerns quantum thermalization and equilibration during coherent quantum
dynamics without coupling to an environment, which appears all but settled.
In the infinitely-connected limit of $\alpha = 0$ it is already well known that
thermalization does not occur~\cite{Sciolla2011, Homrighausen2017}.
Furthermore, numerical works close to this infinitely-connected limit have
already observed indications that thermalization could be prevented at least on
the achievable time scales~\cite{Halimeh2017, Zunkovic2018}.
In order to settle this fundamental question, the understanding of the thermal
equilibrium phases and properties, to which this work contributes, represents a
first key step.
While thermalization corresponds to ensemble equivalence of the thermal
ensemble with the diagonal ensemble, capturing the long-time steady states
during dynamics~\cite{Rigol2012}, it is also not known to which extent such
long-range models exhibit ensemble equivalence on a general level.
This concerns for instance the equivalence of the thermal and microcanonical
ensemble, which is of central importance from the statistical physics point of
view.

\section*{Acknowledgements}

We gratefully acknowledge discussions with K. Schmidt, A. Trombettoni, S.
Ruffo, and A. Silva.

\paragraph{Funding information}
The work of AA, MD and EGL is partly supported by the ERC under grant number
758329 (AGEnTh), by the MIUR Programme FARE (MEPH), and has received funding
from the European Union's Horizon 2020 research and innovation programme under
grant agreement No 817482. This work has been carried out within the activities
of TQT.
This project has received funding from the European Research Council (ERC)
under the European Unions Horizon 2020 research and innovation programme (grant
agreement No. 853443), and MH further acknowledges support by the Deutsche
Forschungsgemeinschaft via the Gottfried Wilhelm Leibniz Prize program.

\bibliography{paper.bib}

\nolinenumbers

\end{document}